\documentclass[12pt]{iopart}
\usepackage{lscape}
\usepackage{iopams}
\usepackage{graphicx}
\RequirePackage{xspace}

\begin{document}
\title[Ab initio calculations of $^{14}$N and $^{15}$N hyperfine structures \ldots]{{\it Ab initio} calculations of  $^{14}$N and $^{15}$N hyperfine structures}
\author{P  J\"onsson$^{1}$, T  Carette$^{2}$, M  Nemouchi$^{3}$  and M  Godefroid$^{2}$ }

\address{$^{1}$  Center for Technology Studies, Malm\"{o} University, 205-06 Malm\"{o}, Sweden }

\address{$^{2}$ Chimie Quantique et Photophysique, CP160/09, Universit\'e Libre de Bruxelles, \\ 
Av. F.D. Roosevelt 50, B-1050 Brussels, Belgium} 

\address{$^{3}$ Laboratoire d'\'Electronique Quantique, Facult\'e de Physique, USTHB, 
BP32, El-Alia, Algiers, Algeria}

\ead{mrgodef@ulb.ac.be}
\begin{abstract}
Hyperfine structure parameters are calculated for the $2p^2 (^3P) 3s \; ^4P_J$, $2p^2 (^3P) 3p \; ^4P^o_J$ and
$2p^2 (^3P) 3p \; ^4D^o_J$ levels, using the {\em ab initio} multiconfiguration 
Hartree-Fock method. The theoretical hyperfine coupling constants are in complete
disagreement with the experimental values of Jennerich {\em et al.}  \cite{Jenetal:06a} deduced from
the analysis of the near-infrared Doppler-free saturated absorption spectra.
\end{abstract}

\submitto{\jpb}

\vfill
\noindent{\hfill \bf \today}

\maketitle


\section{Introduction}

Doppler-free spectra of the near-infrared N~I transitions in the 
$2p^2 (^3P) 3s \; ^4P \rightarrow 2p^2 (^3P) 3p \; ^4P^o$ and
$2p^2 (^3P) 3s \; ^4P \rightarrow 2p^2 (^3P) 3p \; ^4D^o$
multiplets have been recorded by Jennerich {\it et al.}~\cite{Jenetal:06a} using saturated absorption spectroscopy, extending the pioneer work of Cangiano~{\it et al.}~\cite{Canetal:94a} using a similar set-up.
From the analysis of these spectra, Jennerich {\it et al.}~\cite{Jenetal:06a} determined the hyperfine structure 
constants  for the various $J$-values of the three multiplets involved, for both isotopes $^{14}$N and $^{15}$N. 
The isotope shifts in each multiplet have also been measured, revealing a significant \mbox{$J$-dependence} of the shifts. 
These authors recommended a theoretical investigation of the underlying cause of this unexpected phenomenon. 
Similar measurements, using Doppler-free saturated absorption spectroscopy, have been performed in atomic 
Fluorine~\cite{TatAtu:97a}, Chlorine~\cite{TatWal:99a} and Oxygen~\cite{JenTat:00a}.

The present work presents a robust {\em ab initio} theoretical estimation of the relevant hyperfine structure parameters,  
using the ATSP2K package \cite{Froetal:07a}, based on the non relativistic multiconfiguration Hartree-Fock (MCHF) and 
configuration interaction (CI) methods. The minimum theoretical background is presented in section~\ref{hi} for the hyperfine interaction.
Section~\ref{cm} is dedicated to the description of the electron correlation models. 
A refined calculation for the even-parity term is presented in section~\ref{MR2}.
A first comparison between theory and observation appears in 
section~\ref{ce}, revealing unexpected discrepancies. 
Relativistic effects are investigated in section~\ref{re_cor} through the relativistic configuration interaction (RCI) approach, confirming serious problems in the experimental analysis as discussed in section~\ref{concl}. 

\section{The hyperfine interaction}
\label{hi}

The theory underlying MCHF calculations of hyperfine structure parameters can be 
found in \cite{Hib:75b,Jonetal:93a}.
Neglecting the relativistic effects, the diagonal and off-diagonal $A$ and 
$B$ hyperfine interaction constants are expressed in terms of the $J$-independent orbital~($a_{l}$),
spin-dipole~($a_{sd}$), contact~($a_{c}$) and electric quadrupole ($b_{q}$) electronic hyperfine parameters defined by
\begin{eqnarray}
\label{small_a_l}
a_{l}&=&\langle \gamma LSM_{L}M_{S}|\sum_{i=1}^{N}
l^{(1)}_{0}(i)r^{-3}_{i}|\gamma LSM_{L}M_{S}\rangle \; ,  \\
\label{small_a_sd}
a_{sd}&=&\langle \gamma LSM_{L}M_{S}|\sum_{i=1}^{N}
2C^{(2)}_{0}(i)s^{(1)}_{0}(i)r^{-3}_{i}|\gamma LSM_{L}M_{S}\rangle  \; , \\
\label{small_a_c}
a_{c}&=&\langle \gamma LSM_{L}M_{S}|\sum_{i=1}^{N}
2s^{(1)}_{0}(i)r^{-2}_{i}\delta(r_{i})|\gamma LSM_{L}M_{S}\rangle  \; , \\
\label{small_b_hfs_par}
b_{q} &=&\langle \gamma LSM_{L}M_{S}|\sum_{i=1}^{N}
2C^{(2)}_{0}(i)r^{-3}_{i}|\gamma LSM_{L}M_{S}\rangle  \; ,
\end{eqnarray}
and calculated for the magnetic component $M_{L}=L$ and $M_{S}=S$.
The first three parameters (\ref{small_a_l}), (\ref{small_a_sd}) and (\ref{small_a_c}), contribute to the magnetic dipole hyperfine interaction constant through
\begin{equation}
\label{A_constant}
A_J = A_{J}^{orb} + A^{sd}_{J} + A^{c}_{J} \; ,
\end{equation}
with
\begin{eqnarray}
\label{A_orb}
  A_{J}^{orb} & = & G_{\mu}\frac{\mu_{I}}{I}\;a_{l} \; \frac{\langle\,\vec{L}.\vec{J}\,\rangle}{LJ(J+1)}  \; ,\\
 \label{A_sd}
A^{sd}_{J} &= &\frac{1}{2}\,G_{\mu}\,g_{s}\,\frac{\mu_{I}}{I}\,a_{sd} \; \frac{3\,\langle\,\vec{L}.\vec{S}
\,\rangle\,
\langle\,\vec{L}.\vec{J}\,\rangle\,-\,L(L+1)\,\langle\,\vec{S}.\vec{J}\,\rangle}{SL(2L-1)J(J+1)} \; ,\\
\label{A_c}
A_{J}^{c} &=&\frac{1}{6}\,G_{\mu}\,g_{s}\,\frac{\mu_{I}}{I}a_{c} \;
\frac{\langle\,\vec{S}.\vec{J}\,\rangle}{SJ(J+1)} \; ,
\end{eqnarray}
while the last one ($b_q$) constitutes the electronic contribution to the electric quadrupole hyperfine interaction
\begin{equation}
\label{B_constant}
B_{J}=-G_{q}\,Q\,b_{q}\,\frac{6\langle\,\vec{L}.\vec{J}\,\rangle^{2}\,-\,3\langle\,\vec{L}.\vec{J}\,
\rangle\,-\,2L(L+1)J(J+1)}{L(2L-1)(J+1)(2J+3)} \; .
\end{equation}
Expressing the electronic parameters $a_l$, $a_{sd}$ and $a_c$ in atomic units ($a_0^{-3}$) and $\mu_I$ in nuclear magnetons ($\mu_N$), the magnetic dipole hyperfine structure constant $A$  is calculated in units of frequency (MHz) by using $G_\mu = 95.41067$. Similarly, the  electric quadrupole hyperfine structure constant $B$ is expressed in MHz  when adopting atomic units ($a_0^{-3}$) for $b_q$, barns for $Q$ and $G_q = 234.96475$.

The electronic parameter governing the mass isotope shift of an atomic energy level is the
$S_{sms} $ parameter~\cite{Godetal:01a} defined by 
\begin{equation}
\label{sms}
S_{sms} = 
- \langle  \gamma LSM_{L}M_{S} | \sum_{i<j}^N 
\vec{\nabla}_i \cdot \vec{\nabla}_j | \; \gamma LSM_{L}M_{S}  \rangle \; . 
\end{equation}

\section{Correlation models}
\label{cm}

The multiconfiguration Hartree-Fock (MCHF) variational approach consists in optimizing the 
one-electron functions spanning a configuration space and the mixing coefficients of the interacting configuration state functions (CSF)~\cite{FBJ:97a} for describing a given term
\begin{equation}
\label{eq:SOC}
\Psi(\gamma LS M_L M_S) = \sum_{i} c_i \Phi(\gamma_i LSM_L M_S).
\end{equation} 
Efficient MCHF expansions are often built by allowing 
single and double excitations from a multireference set (SD-MR-MCHF). As far as hyperfine structures are concerned,
successful applications of this method are found for light elements such as Li-like ions~\cite{Godetal:01a},  
Be~I, B~I, C~II and C~I~\cite{JonFro:93a,Jonetal:96a}, N~I~\cite{Godetal:97b}, O~I~\cite{Godetal:97b,JonGod:00a} or
Na~I~\cite{Jonetal:96b}.  \\


For the even parity term $2p^23s~^4P$, the configuration expansion is generated from single and double (SD) excitations from the 
\begin{equation}
\{ 2s^2 2p^2 3s, \; 2p^4 3s ,  \; 2s 2p^4 \}
\end{equation}
multireference (MR) to increasing active sets (AS) of orbitals that are denoted by specifying the number of orbitals for each $l$-symmetry.  This multireference set  captures the
dominant correlation effects through a physical good ``zero-order'' wave function. We include only 
configuration state functions ($\Phi(\gamma_i LSM_L M_S)$)  that interact with the multireference, adopting the reversed orbital order, {\it i.e.} coupling sequentially the subshells by decreasing $n$ and $l$. This technique reduces substancially the size of the MCHF expansions while keeping the dominant correlation contributions~\cite{Jonetal:96a,Caretal:10a}.
With  the largest set of orbitals ($10s9p8d6f3g$) optimized through these calculations, 
the effect of higher excitations is investigated through configuration interaction (CI) calculations
using the configuration state function set obtained by adding to the original SD-MR CSF list, the
triple  and quadruple excitations from the same multireference  to smaller orbital active sets 
(up to $6s5p4d3f$).
For generating these lists, some limitations have been introduced by imposing the restriction that 
there should be at least 5 orbitals with $n \leq 3 $ in the CSFs produced.
 The merging of the original SD-MR and the TQ-MR CSF lists is
noted by the union ($\cup$) symbol. \\

For the odd  parity states $2p^23p~^4P^o$ and $2p^23p~^4D^o$, a similar strategy is adopted, 
using the following multireference set
\begin{equation}
\{ 2s^22p^23p, \; 2p^43p , \; 2s2p^23p3d \} \; .
\end{equation}

The total energies ($E$), the specific mass shift ($S_{sms} $)  and the hyperfine interactions parameters $\{a_l, a_{sd}, a_c, b_q \}$ defined in section~\ref{hi},
are reported in Tables~\ref{tab:hfs_para_even}, \ref{tab:hfs_para_odd_4P}  and \ref{tab:hfs_para_odd_4D},
respectively. These tables illustrate the smooth convergence of the various parameters while 
improving the correlation model by increasing the size of the orbital active set. 
For the even parity term $2p^23s~^4P$ (Table~\ref{tab:hfs_para_even}), triple and quadruple excitations from the multireference affect the
 $a_l$  and  $a_{sd}$ parameters at the level of 2\% while the contact term $a_c$ is much more  sensitive~(33\%).
  For the odd parity $2p^23p~^4P^o$ and $2p^23p~^4D^o$ terms (Tables~\ref{tab:hfs_para_odd_4P}  and \ref{tab:hfs_para_odd_4D}),
 all hyperfine parameters  reach a high degree of convergence, except the contact parameter that is strongly affected (around 30\%) by TQ excitations.
 For the three terms, the last layer added at the SDTQ-CI level of approximation brings a negligible increment, except for the contact contributions (3\%). The convergence patterns of the contact contributions, as well as the sensitivity for TQ excitations, are similar to the one found for the $2p^33s~^5S^o$ term in O I \cite{Godetal:97b}.
 
\begin{table} [!h]
\caption{\label{tab:hfs_para_even} SD-MR-MCHF and CI total energies, specific mass shift and hyperfine interaction parameters (all in atomic units) as a function of the orbital active set, for $2p^23s~^4P$.
NCSF is the number of CSFs in the wavefunction expansion.}
\begin{indented}
\item[]
\begin{tabular}{lrcccccc}    
\br
AS               & NCSF      &         $E$   &  $S_{sms} $     &  $a_l$ & $a_{sd}$  & $a_c$   & $b_q$   \\ \hline 
\multicolumn{8}{c}{SD-MR-MCHF} \\ \hline
HF               & 1      & $-$54.032~303 & $-$2.5647  & 3.7993 & 0.7599  & 3.6055  & 1.5197  \\
3s2p1d           & 149    & $-$54.116~873 & $-$2.3529  & 3.5643 & 0.7310  & 8.4493  & 1.2085  \\
4s3p3d1f         & 652    & $-$54.165~015 & $-$1.3937  & 3.7020 & 0.7999  & 9.2913  & 1.2704  \\
5s4p3d2f1g       & 1626   & $-$54.189~786 & $-$1.4252  & 3.6990 & 0.7837  & 7.3833  & 1.2072  \\
6s5p4d3f2g       & 3082   & $-$54.196~747 & $-$1.4254  & 3.6862 & 0.7713  & 8.7173  & 1.2108  \\
7s6p5d4f3g       & 5020   & $-$54.199~972 & $-$1.4238  & 3.6869 & 0.7723  & 8.1898  & 1.2281  \\
8s7p6d5f3g       & 7113   & $-$54.201~188 & $-$1.4238  & 3.6879 & 0.7751  & 8.0753  & 1.2248  \\
9s8p7d6f3g       & 9572   & $-$54.201~746 & $-$1.4237  & 3.6878 & 0.7752  & 8.0941  & 1.2208  \\
10s9p8d6f3g      & 11728  & $-$54.201~953 & $-$1.4234  & 3.6884 & 0.7743  & 8.1090  & 1.2231  \\ 
\hline
\multicolumn{8}{c}{SDTQ-CI} \\ \hline
$\cup$~4s3p2d1f   & 40685  & $-$54.203~791 & $-$1.4282  & 3.6412 & 0.7668  & 9.7166  & 1.1502  \\
$\cup$~5s4p3d2f   & 106472 & $-$54.204~677 & $-$1.4427  & 3.6170 & 0.7624  & 10.511  & 1.1156  \\
$\cup$~6s5p4d3f   & 210533 & $-$54.204~967 & $-$1.4470  & 3.6090 & 0.7609  & 10.780  & 1.1049  \\
\br 
\end{tabular}
\end{indented}
\end{table}

\begin{table} [!h]
\caption{\label{tab:hfs_para_odd_4P} SD-MR-MCHF and CI total energies, specific mass shift and hyperfine interaction parameters (all in atomic units) as a function of the orbital active set, for $2p^23p~^4P^o$.
NCSF is the number of CSFs in the wavefunction expansion.}
\begin{indented}
\item[]
\begin{tabular}{lrcccccc}    
\br
AS               & NCSF      &         $E$   &  $S_{sms} $     &  $a_l$ & $a_{sd}$  & $a_c$   & $b_q$   \\ \hline
\multicolumn{8}{c}{SD-MR-MCHF} \\ \hline
HF               & 1      & $-$53.984~055 & $-$2.5786  & 1.9609 & $-$0.3749 & 0.0000  & $-$0.7498 \\
3s2p1d           & 583    & $-$54.062~052 & $-$2.2945  & 1.9466 & $-$0.3787 & 5.1320  & $-$0.7123 \\
4s3p2d1f         & 3879   & $-$54.112~044 & $-$1.2732  & 1.9752 & $-$0.4086 & 0.5548  & $-$0.6912 \\
5s4p3d2f1g       & 10078  & $-$54.134~423 & $-$1.3540  & 1.9653 & $-$0.3884 & 0.3990  & $-$0.6708 \\
6s5p4d3f2g       & 19200  & $-$54.143~169 & $-$1.3318  & 1.9673 & $-$0.3839 & 1.1424  & $-$0.6841 \\
7s6p5d4f3g       & 31245  & $-$54.147~455 & $-$1.3170  & 1.9709 & $-$0.3835 & 0.7422  & $-$0.6879 \\
8s7p6d5f3g       & 44096  & $-$54.148~983 & $-$1.3187  & 1.9705 & $-$0.3837 & 0.6826  & $-$0.6887 \\
9s8p7d6f3g       & 59110  & $-$54.149~672 & $-$1.3185  & 1.9702 & $-$0.3850 & 0.7044  & $-$0.6864 \\
10s9p8d6f3g      & 72070  & $-$54.149~921 & $-$1.3188  & 1.9704 & $-$0.3843 & 0.7449  & $-$0.6849 \\ 
\hline
\multicolumn{8}{c}{SDTQ-CI} \\ \hline
$\cup$~4s3p2d1f   & 124029 & $-$54.150~179 & $-$1.3136  & 1.9710 & $-$0.3844 & 0.8878  & $-$0.6848 \\
$\cup$~5s4p3d2f   & 252690 & $-$54.150~330 & $-$1.3139  & 1.9714 & $-$0.3844 & 0.9168  & $-$0.6845 \\
$\cup$~6s5p4d3f   & 459494 & $-$54.150~406 & $-$1.3138  & 1.9717  &$-$0.3843 & 0.9429  & $-$0.6845 \\  
\br 
\end{tabular}
\end{indented}
\end{table}

\begin{table} [!h]
\caption{\label{tab:hfs_para_odd_4D} SD-MR-MCHF and CI total energies, specific mass shift and hyperfine interaction parameters (all in atomic units) as a function of the orbital active set, for $2p^23p~^4D^o$.
NCSF is the number of CSFs in the wavefunction expansion.}
\begin{indented}
\item[]
\begin{tabular}{lrcccccc}    
\br
AS               & NCSF      &         $E$   &  $S_{sms} $     &  $a_l$ & $a_{sd}$  & $a_c$   & $b_q$   \\ \hline 
\multicolumn{8}{c}{SD-MR-MCHF} \\ \hline
HF               & 1      & $-$53.987~051 & $-$2.5789  & 3.9263 &  0.7462 &  0.0000 &  1.4926 \\
3s2p1d           & 720    & $-$54.064~819 & $-$2.2937  & 3.8979 &  0.7539 &  5.1069 &  1.4194 \\
4s3p2d1f         & 4747   & $-$54.114~835 & $-$1.2729  & 3.9574 &  0.8137 &  0.5229 &  1.3775 \\
5s4p3d2f1g       & 12197  & $-$54.137~237 & $-$1.3541  & 3.9575 &  0.7769 &  0.3613 &  1.3421 \\
6s5p4d3f2g       & 23072  & $-$54.146~073 & $-$1.3259  & 3.9674 &  0.7698 &  1.1034 &  1.3472 \\
7s6p5d4f3g       & 37372  & $-$54.150~182 & $-$1.3163  & 3.9688 &  0.7661 &  0.6698 &  1.3768 \\
8s7p6d5f3g       & 52558  & $-$54.151~947 & $-$1.3186  & 3.9688 &  0.7658 &  0.6258 &  1.3800 \\
9s8p7d6f3g       & 70266  & $-$54.152~646 & $-$1.3188  & 3.9678 &  0.7692 &  0.6472 &  1.3730 \\
10s9p8d6f3g      & 85436  & $-$54.152~903 & $-$1.3191  & 3.9682 &  0.7688 &  0.6864 &  1.3686 \\
\hline
\multicolumn{8}{c}{SDTQ-CI} \\ \hline
$\cup$~4s3p2d1f   & 156938 & $-$54.153~165 & $-$1.3139  & 3.9695 &  0.7691 &  0.8307 &  1.3684 \\ 
$\cup$~5s4p3d2f   & 334710 & $-$54.153~318 & $-$1.3142  & 3.9706 &  0.7690 &  0.8600 &  1.3678 \\ 
$\cup$~6s5p4d3f   & 620614 & $-$54.153~406 & $-$1.3138  & 3.9711 &  0.7689 &  0.8848 &  1.3677 \\ 
\br 
\end{tabular}
\end{indented}
\end{table}

\section{On a larger Multireference set for $2p^2(^3P)3s \; ^4P$}
\label{MR2}

We investigate the reliability of the 
theoretical parameters by extending the multireference set. Amongst the three terms considered, 
we focus on the even parity $2p^2 (^3P) 3s \; ^4P $ one, realizing that it is one for 
which the hyperfine interaction parameters $A_J$ are the most sensitive to TQ excitations due to the large contribution of the contact term. The following multireference (MR')
\begin{equation}
\{ 2s^2 2p^2 3s, \; 2p^4 3s ,  \; 2s 2p^4 , \; 2s 2p^3 3p , \; 2s 2p^2 3s 3d \}
\end{equation}
is selected, after a detailed analysis of the eigenvector weights obtained with the first approach. 
The results are reported in Table~\ref{tab:hfs_para_even_MR2}. 
The comparison of the last lines of
Table~\ref{tab:hfs_para_even}~(MR) and Table~\ref{tab:hfs_para_even_MR2}~(MR') illustrates the global stability of the hyperfine parameters, 
the largest variation (3.6\%) being observed for the $a_c$ contact contribution. 

\begin{table} [!h]
\caption{\label{tab:hfs_para_even_MR2} SD-MR'-MCHF and CI total energies, specific mass shift and hyperfine interaction parameters (all in atomic units) as a function of the orbital active set, for $2p^23s~^4P$.
NCSF is the number of CSFs in the wavefunction expansion.}
\begin{indented}
\item[]
\begin{tabular}{lrcccccc}    
\br
AS               & NCSF      &         $E$   &  $S_{sms} $     &  $a_l$ & $a_{sd}$  & $a_c$   & $b_q$   \\ \hline
\multicolumn{8}{c}{SD-MR'-MCHF} \\ \hline
HF               & 1      & $-$54.032~303 & $-$2.5647  & 3.7993 & 0.7599  & 3.6055  & 1.5197  \\
3s2p1d           & 461    & $-$54.117~540 & $-$2.3595  & 3.5468 & 0.7276  &  9.0794  & 1.1821 \\
4s3p2d1f         & 2543   & $-$54.168~450 & $-$1.3886  & 3.6009 & 0.7798  &  5.8010  & 1.1433 \\
5s4p3d2f1g       & 6447   & $-$54.192~356 & $-$1.4432  & 3.6402 & 0.7761  &  9.6255  & 1.1132 \\
6s5p4d3f2g       & 12210  & $-$54.199~795 & $-$1.4511  & 3.6101 & 0.7584  & 11.3959  & 1.0968 \\
7s6p5d4f3g       & 19832  & $-$54.203~174 & $-$1.4498  & 3.6078 & 0.7589  & 10.8596  & 1.1146 \\
8s7p6d5f3g       & 28014  & $-$54.204~436 & $-$1.4505  & 3.6080 & 0.7615  & 10.7556  & 1.1106 \\
9s8p7d6f3g       & 37595  & $-$54.205~011 & $-$1.4509  & 3.6075 & 0.7615  & 10.7965  & 1.1068 \\
10s9p8d6f3g      & 45958  & $-$54.205~221 & $-$1.4505  & 3.6084 & 0.7607  & 10.7996  & 1.1094 \\ 
\\ \hline 
\multicolumn{8}{c}{SDTQ-CI} \\ \hline
$\cup$~4s3p2d1f   & 93412  & $-$54.205~400 & $-$1.4538  & 3.5997 & 0.7592  & 11.0883  & 1.0953 \\
$\cup$~5s4p3d2f   & 208694 & $-$54.205~456 & $-$1.4540  & 3.5977 & 0.7588  & 11.1538  & 1.0922 \\
$\cup$~6s5p4d3f   & 393284 & $-$54.205~469 & $-$1.4543  & 3.5972 & 0.7588  & 11.1705  & 1.0913 \\
\br
\end{tabular}
\end{indented}
\end{table}

\section{Comparison with experiments}
\label{ce}
The hyperfine constants $A_J$ and $B_J$ are estimated for both isotopes $^{14}$N and $^{15}$N 
from the hyperfine structure parameters calculated with the most elaborate correlation models 
(last lines of Tables~\ref{tab:hfs_para_odd_4P}, \ref{tab:hfs_para_odd_4D} and \ref{tab:hfs_para_even_MR2}),
using the nuclear data taken from Stone~\cite{Sto:05a} and  summarized in Table~\ref{tab:nuclear_data}. 
\begin{table} [!h]
\caption{\label{tab:nuclear_data} Nuclear data for $^{14}$N and $^{15}$N~\cite{Sto:05a}.}
\begin{indented}
\item[]
\begin{tabular}{cccc}
\br
isotope & $I$     &   $\mu_I$~(nm)   &  $Q$~(b)   \\
\hline
$^{14}$N & $1$                     & $+$0.40376100(6)  & $+$0.02001(10)\\
$^{15}$N & $\frac{1}{2}$           & $-$0.28318884(5)  & 0.0 \\
\br 
\end{tabular}
\end{indented}
\end{table}
From equations~(\ref{A_orb})-(\ref{A_c}), one realizes that the ratio between the magnetic hyperfine 
constants characterizing a given $J$-level of the two isotopes should be
\begin{equation}
A_{J}(^{15}\mbox{N})/A_{J}(^{14}\mbox{N})= 
\frac{\mu_{I}(^{15}\mbox{N}) I(^{14}\mbox{N})}{\mu_{I}(^{14}\mbox{N}) I(^{15}\mbox{N})} =-1.4028\, .
\end{equation}
These non relativistic (NR) theoretical $A_J$ and $B_J$ hyperfine constants are reported in Table~\ref{comp_theory_exp}  and compared with those derived by
Jennerich {\it et al.}~\cite{Jenetal:06a} using saturated absorption spectroscopy~\footnote{As discussed in~\cite{Jenetal:06a}, the hyperfine constants derived by Cangiano {\it et al.}~\cite{Canetal:94a} for $^4P_J$ and $^4P^o_J$ agree qualitatively with those of Jennerich {\it et al.}~\cite{Jenetal:06a}.}.
\begin{table} [!h]
\caption{\label{comp_theory_exp} Hyperfine constant comparison between observation~\cite{Jenetal:06a} and theory (present work) :
non relativistic (NR) {\em ab initio} and relativistically corrected ($+$RC) results. All values are in MHz. }
\scriptsize
\begin{tabular}{c|ccc|cccccc}
\br
 & \multicolumn{3}{c|}{$^{15}$N} & \multicolumn{6}{c}{$^{14}$N} \\
Levels & exp.~\cite{Jenetal:06a} & \multicolumn{2}{c|}{theory}  &\multicolumn{2}{c}{exp.~\cite{Jenetal:06a}} & \multicolumn{4}{c}{theory}  \\
&  & NR & $+$RC & &   & \multicolumn{2}{c}{NR}  & \multicolumn{2}{c}{$+$RC} \\
& $A$ & $A$ & $A$ & $A$ & $B$ & $A$ & $B$ & $A$ & $B$ \\
\hline
$^4P_{1/2}$ &$\left\{\begin{array}{c}
                        +103.4(14)\\
                        -153.1(23)^{a}\\ 
                        \end{array} \right. $ &  $-$139.85 & $-$140.56 &$\left\{\begin{array}{c}
                        -69.76(90) \\
                        +112.3(13)^{a}\\ 
                        \end{array} \right.$ & 0.0 &  99.70  & 0.0 & 100.21& 0.0 \\
$^4P_{3/2}$ & $-$47.93(48) &  $-$88.29 & $-$87.62 & 35.52(44) & $-$0.98(48) &  62.94  &  4.10 & 62.46 & 4.10\\
$^4P_{5/2}$ & $-$90.71(71) &  $-$174.75& $-$175.12& 64.76(42) & $-$3.9(10) &  124.58 & $-$5.13 & 124.84 & $-$5.12 \\
&&&&&&&&& \\
$^4P^o_{1/2}$ & 167.1(13) & 75.24 & 73.29  &$-$133.2(22) & 0.0 & $-$53.64 & 0.0 & $-$52.25 & 0.0  \\
$^4P^o_{3/2}$ & 70.0(12) & $-$68.15 & $-$71.60 & $-$48.56(74) & 8.69(87) & 48.58 & $-$2.58 & 51.04 & $-$2.95 \\
$^4P^o_{5/2}$ & 46.20(74) & $-$41.11 & $-$46.52 & $-$32.83(44) & 5.0(11) & 29.30 & 3.22 & 33.16  &  2.57 \\
&&&&&&&&& \\
$^4D^o_{1/2}$ &$\left\{\begin{array}{c}
                       +153.1(23)\\
                       -103.4(14)^{a}\\ 
                        \end{array} \right. $ & $-$106.89 & $-$104.02 & $\left\{\begin{array}{c}
                        -112.3(13)\\
                        +69.76(90)^{a}\\ 
                        \end{array} \right. $ & 0.0 & 76.20 & 0.0 & 74.15 & 0.0 \\
$^4D^o_{3/2}$ & 92.4(17) & $-$49.14 & $-$44.49 & $-$64.41(79) & 10.46(88) & 35.03 & 0.0 & 31.71 & 0.30  \\
$^4D^o_{5/2}$ & 41.5(14) & $-$56.74 & $-$51.57 & $-$28.19(62) & $-$0.2(15) & 40.45 & $-$2.30 & 36.76 & $-$1.69  \\
$^4D^o_{7/2}$ & $-$9.35(55) & $-$77.76 & $-$78.04 & 6.31(72) & $-$12.6(13) & 55.43 & $-$6.43 & 55.63 & $-$6.44  \\
\br 
\multicolumn{10}{l}{$^{a}$~Second proposition of Jennerich \emph{et al} (see text)}
\end{tabular}
\end{table}
The comparison reveals huge discrepancies between theory and observation. 
Inconsistencies appear not only in the magnitude of the parameters, but even in the relative signs of the parameters for the 
different $J$-levels arising from the same term. 

\clearpage

\noindent
Rewriting equations~(\ref{small_a_l})-(\ref{small_b_hfs_par}) as
\begin{eqnarray}
\label{A_orb_secnd_form}
  A_{J}^{orb} & \equiv &   G_{\mu}\frac{\mu_{I}}{I}\;a_{l} \; K^{orb}_J \; ,  \\
\label{A_sd_secnd_form}
A^{sd}_{J} & \equiv &  \frac{1}{2}\,G_{\mu}\,g_{s}\,\frac{\mu_{I}}{I}\,a_{sd} \; K^{sd}_J \; , \\
\label{A_c_secnd_form}
A_{J}^{c} & \equiv &\frac{1}{6}\,G_{\mu}\,g_{s}\,\frac{\mu_{I}}{I}a_{c} \; K^c_J \; , \\
\label{B_secnd_}
B_{J} & \equiv & -G_{q}\,Q\,b_{q} \; K '_J  \; ,
\end{eqnarray}
the numerical factors $K^{orb}_J$, $K^{sd}_J$, $K^c_J$ and $K '_J$ are calculated from the expectation 
values of the angular momentum scalar products and reported in Table~\ref{tab:J_dependence} for each $J$-values of the 
(odd and even) $^{4}P$  and $^{4}D$ terms.
\begin{table} [!h]
\caption{\label{tab:J_dependence} $J$-dependence of the hyperfine contributions to $A_J$ and $B_J$  constants for $^{4}P$ and $^{4}D$ terms.}
\begin{indented}
\item[]
\begin{tabular}{c|rrrr|rrrr}
\br
$J$     &   $K_J^{orb}$    &  $K_J^{sd}$  &   $K_J^c$    &  $K'_J$ &   $K_J^{orb}$    &  $K_J^{sd}$  &   $K_J^c$    &  $K'_J$ \\
\multicolumn{1}{c|}{\ } &
\multicolumn{4}{c|}{$^4P$} & \multicolumn{4}{c}{$^{4}D$} \\
\hline
$1/2$  &   $-\frac{2}{3}$   &   $\frac{10}{9}$& $ \frac{10}{27}$   &  0
& 1   &   $-\frac{7}{3}$   & $  -\frac{2}{9}$  &  0\\
$3/2$   &  $ \frac{4}{15}$&  $-\frac{68}{45} $  &  $ \frac{22}{135}$  & $-\frac{4}{5}$  
& $\frac{2}{5}$   & $ -\frac{14}{15}$   & $  \frac{2}{45} $ & 0  \\
$5/2$   &   $\frac{2}{5}$   & $ \frac{2}{5}$  & $  \frac{2}{15} $ & 1 
& $ \frac{11}{35} $  & $ -\frac{37}{105} $  & $   -\frac{26}{315} $ &$ \frac{9}{25}$ \\
$7/2$  &   &  &  &  
&  $\frac{2}{7}$   &   $\frac{2}{7}$   &    $\frac{2}{21}$  & 1 \\
\br
\end{tabular}
\end{indented}
\end{table}
The consistency of the theoretical values calculated for the different $J$-values within a given 
$LS$ term with the relative weights of the different contributions 
making the total hyperfine constant, is obviously satisfied, by construction. However there is no such constraint in the experimental analysis and one can show that no physical set of underlying $\{a_l, a_{sd}, a_c \}$ parameters fits the experimental $A_J$-values with the numerical factors
of Table~\ref{tab:J_dependence}.

Some ambiguity was pointed out in the line assignment of the spectra of the  
$^4P_{1/2} \rightarrow \; ^4D^o_{1/2}$ transition for both  isotopes. 
 Due to this identification problem, Jennerich  \textit{et al}~\cite{Jenetal:06a} determined two possible values 
 for each of the hyperfine constants of the $^4P_{1/2}$ and $^4D^o_{1/2}$ states, both reported in Table~\ref{comp_theory_exp}. On the basis of crossover intensity arguments, they gave their preference to the first set. Looking to our theoretical values, we claim that their choice was definitely not the good one. 
Another surprise appears: they deduced a large $B$ value for $^4D^o_{3/2}$~$^{14}$N that contrasts with the zero numerical factor $K'_{3/2}$ of Table~\ref{tab:J_dependence}.

\section{Relativistic corrections}
\label{re_cor}

 Relativistic effects influence atomic wave functions in basically two ways: through
contraction of radial orbitals and through $LS$ term mixing. Contraction effects may be large for high~$Z$, but remain relatively unimportant in light elements like Nitrogen~\cite{LinRos:74a}. However, term mixing may affect the wave function and the 
computed hyperfine interaction constants in significant ways, especially when fine-structure levels belonging to different terms are closely spaced.
To investigate term mixing we first perform reference MCHF calculations for configuration expansions generated by SD-excitations from, respectively, $\{2s^22p^23s, 2p^43s\}$ and $\{2s^22p^23p, 2p^43p\}$ to active 
sets $(4s3p2d1f)$. The resulting non-relativistic radial orbitals are converted to Dirac spinors using the Pauli
approximation~\cite{ArmFen:74a}
\begin{eqnarray}
P(n \kappa ; r ) =  P^{MCHF}(nl;r) \; ,\\ \nonumber
Q(n \kappa ; r ) = \frac{\alpha}{2} \left( \frac{d}{dr} + \frac{\kappa}{r} \right) P(n \kappa ; r) \; .
\end{eqnarray}
Here $\alpha$ is the fine structure constant and $\kappa$ the relativistic quantum number
\begin{equation}
\kappa = (j+1/2) \eta 
\hspace*{0.5cm} \mbox{when} \hspace*{0.5cm} l = j + \eta /2 , \hspace*{0.5cm} \eta = \pm 1 \; .
\end{equation}
 This is followed by
relativistic configuration interaction (RCI) calculations for configuration expansions generated by SD-excitations from the multireference sets above to the Dirac spinors (see ref. \cite{Jonetal:07a,Jonetal:96c} for details about the conversion of radial orbitals and the relativistic computer codes). In relativistic theory only $J$ is a good quantum number and the relativistic configuration expansions account for $LS$ term mixing. The hyperfine interaction constants obtained from the MCHF and RCI wave functions are displayed
in Table~\ref{tab:RCI}. One observes that the relativistic effects are far from negligible, changing the hyperfine parameters in many cases by more than 10~\%. Assuming that the differences between the relativistic and non-relativistic values of the hyperfine constants in the limited calculations with active sets $(4s3p2d1f)$ are representative of the true differences, we add the former 
to the non-relativistic results in Table~\ref{comp_theory_exp} to obtain the final relativistically corrected values (in column ``$+$RC'') of the hyperfine interaction constants.  
It is clear that relativistic effects cannot explain the huge global theory-observation conflict. 

\begin{table}[h]
\caption{\label{tab:RCI}Hyperfine interaction constants from matching RCI and MCHF calculations. The relativistic orbitals were obtained from the
non-relativistic ones in the Pauli approximation. All values are in MHz.}
	\centering
		\begin{tabular}{l|ccc|cccccc} \br
		            & \multicolumn{3}{c|}{$^{15}$N}  & \multicolumn{6}{c}{$^{14}$N}             \\
		            & \multicolumn{1}{c}{RCI} & \multicolumn{1}{c}{MCHF} & \multicolumn{1}{c|}{Difference} 
		            & \multicolumn{2}{c}{RCI} & \multicolumn{2}{c}{MCHF} & \multicolumn{2}{c}{Difference}\\
			Levels    &  $A$          &   $A$        &  $\Delta A$ &     $A$  &  $B$     &   $A$   &  $B$   &  $\Delta A$   &   $\Delta B$   \\ \hline
			                &               &              &             &         &          &         &        &               &                 \\
$^4P_{1/2}$     &  $-$79.79     &    $-$79.08  &  $-$0.71    & 56.88   &  0.0     &  56.37  &  0.0   &  0.51         &    0.0     \\
$^4P_{3/2}$     &  $-$59.96     &    $-$60.63  &   0.67      & 42.74   &  5.06    &  43.22  & 5.06   & $-$0.48       &    0.0     \\
$^4P_{5/2}$     & $-$158.07     &   $-$157.70  &  $-$0.37    & 112.68  & $-$6.32  & 112.42  & $-$6.33&  0.26         &    0.01     \\
                &               &              &             &         &          &         &        &               &                \\
$^4P^o_{1/2}$   &   84.01       &       85.96  &  $-$1.95    & $-$59.89  &  0.0   & $-$61.28&  0.0   &  1.39         &    0.0      \\ 
$^4P^o_{3/2}$   &  $-$69.63     &    $-$66.18  &  $-$3.45    &  49.64  & $-$2.97  &  47.18  & $-$2.60&  2.46         &   $-$0.37      \\
$^4P^o_{5/2}$   &  $-$42.81     &    $-$37.40  &  $-$5.41    &  30.52  &  2.60    &  26.66  &  3.25  &  3.86         &   $-$0.65      \\
                &               &              &             &         &          &         &        &               &                 \\
$^4D^o_{1/2}$   & $-$102.87     &   $-$105.74  &   2.87      &  73.33  & 0.0      & 75.38   &  0.0   & $-$2.05       &    0.0      \\
$^4D^o_{3/2}$   &  $-$40.95     &    $-$45.60  &   4.65      &  29.19  & 0.30     & 32.51   &  0.0   & $-$3.32       &    0.30      \\
$^4D^o_{5/2}$   &  $-$48.61     &    $-$53.78  &   5.17      &  34.65  & $-$1.70  & 38.34   & $-$2.31& $-$3.69       &    0.61      \\
$^4D^o_{7/2}$   &  $-$76.34     &    $-$76.06  & $-$0.28     &  54.42  & $-$6.48  & 54.22   & $-$6.47&  0.2          &   $-$0.01      \\ \br
		\end{tabular}
\end{table}

\section{Conclusion}
\label{concl}

The strong disagreement between theory and observation \cite{Jenetal:06a} is really disconcerting. Hyperfine parameters have indeed been estimated {\it ab initio} successfully using similar methods for different atomic systems such as Carbon, Nitrogen, Oxygen or Sodium, as described in the introduction. The present calculated hyperfine structure constants disagree so strongly with experiments in comparison of the achieved theoretical convergence of the hyperfine parameters that we presently cast doubt on Cangiano~{\em et al}'s \cite{Canetal:94a} and Jennerich {\em et al}'s~\cite{Jenetal:06a} analysis. We therefore encourage further experimental spectroscopic studies and/or reinterpretation of the near-infrared spectra. Last but not least, the extracted isotope shifts from the same spectra  revealed a significant unexpected $J$-dependence of the specific mass shifts in both multiplets. However, the authors themselves \cite{Jenetal:06a} observed that the experimental isotope shift values are critically dependent on the correct interpretation of the hyperfine structures of the  $^{14}$N and  $^{15}$N spectra. The present questioning on the experimental determination of the hyperfine parameters could also be relevant  in their isotope shift discussion.

\ack

TC has a PhD fellowship from the ``Fonds pour la formation \`a la Recherche dans l'Industrie et dans l'Agriculture'' of Belgium (Boursier F.R.S. - FNRS). PJ acknowledges support from the Swedish research council.
The Fonds de la Recherche Scientifique de Belgique FRFC Convention)
and the Communaut\'e fran\c{c}aise de Belgique (Actions de Recherche
Concert\'ees) are acknowledged for their financial support. 

%
%

\section*{References}

\end{document}